\documentclass[a4paper]{jpconf}
\usepackage{graphicx}
\usepackage{amssymb}

\usepackage{tikz}
\usetikzlibrary{arrows,decorations.pathmorphing,backgrounds,positioning,fit,petri}

\begin{document}
\title{Open Quantum Walks: a short introduction}

\author{Ilya Sinayskiy and Francesco Petruccione}

\address{National Institute for Theoretical Physics and Quantum Research Group, School of Chemistry and Physics, University of KwaZulu-Natal, Durban, South Africa}

\ead{sinayskiy@ukzn.ac.za, petruccione@ukzn.ac.za}

\begin{abstract}
The concept of open quantum walks (OQW), quantum walks exclusively driven by the interaction with the external environment, is reviewed. OQWs are formulated as discrete completely positive maps on graphs. 
The basic properties of OQWs are summarised and new examples of OQWs on $\mathbb{Z}$ and their simulation by means of quantum trajectories are presented.
\end{abstract}

\section{Introduction}
The mathematical concept of classical random walks (CRW) \cite{pt} has found wide application in various fields of modern science, i.e. physics \cite{rwp}, computer science \cite{rwcs}, economy \cite{rwe} and biology \cite{rwb}. In this case, the trajectory of the walker is fully determined by the transition matrix of the underlying graph. A quantum extension of this concept was also suggested \cite{aharonov, kempe}. In this case the trajectory of the walker is defined not only by the transition matrix but also by some inner state of the quantum walker, e.g., spin or polarisation. Unitary quantum walks are one of the ways to implement quantum computing algorithms \cite{qwqc1,qwqc2}. Although the practical implementation of any complex quantum system is very difficult due to unavoidable interaction with an environment \cite{toqs}, successful realisations of the unitary quantum walks have been reported. A more adequate description of the experimental realisations of the unitary quantum walks can be obtained by including the effects of decoherence and dissipation.
Usually, in these approaches, the effects of decoherence and dissipation are small and the influence the dynamics of the quantum walker needs to be minimised or eliminated.

Recently, the formalism of discrete-time open quantum walks (OQW) was introduced \cite{pla, JSP}. In an OQWs the quantum walker is exclusively driven by the interaction with an environment. The formalism of OQWs rests upon the implementation of  completely positive maps. OQWs can be used to perform efficient dissipative quantum computation and state engineering \cite{pla, QIP}. The properties of the asymptotic probability distribution were analysed as well \cite{clt1,clt2, PhysSc}. 

The paper has the following structure. In Section 2 we introduce the OQW formalism. In Section 3 we present a simple example of an OQWs on a two-node graph and of an OQWs on $\mathbb{Z}$. In Section 4 we demonstrate the simulation of OQWs by the means of quantum trajectories. In Section 5 the \textit{realisation procedure} to obtain OQWs from unitary dynamics is presented. In Section 6 we show the connection of OQWs to unitary quantum walks and classical random walks. In Section 7 results of the limit theorems are presented. In Section 8 the application of OQW to dissipative quantum computing is described and in Section 9 we conclude.
 
\section{Open Quantum Walks}

OQWs are introduced on a set of vertices $\cal{V}$ with oriented edges $\{(i,j)\,;\ i,j\in\cal{V}\}$. The number of nodes is considered to be finite or countable infinite. The space of states corresponding to the dynamics on the graph specified by the set of nodes $\cal{V}$ will be denoted by  $\cal{K}=\mathbb{C}^\mathcal{V}$. If $\cal{V}$ is an infinite countable set, the space of states $\cal{K}$ will be any separable Hilbert space with an orthonormal basis ${(| i\rangle)}_{i\in\cal{V}}$ indexed by $\cal{V}$. The internal degrees of freedom of the quantum walker, e.g. the spin, polarisation, angular momenta or $n$-energy levels, will be described by a separable Hilbert space $\cal{H}$ attached to each vertex of the graph. Any state of the quantum walker will be described on the direct product of the Hilbert spaces $\cal{H}\otimes \cal{K}$.

\begin{figure}
\begin{center}
\begin{tikzpicture}
[place/.style={circle,draw=black!50,fill=white!20,thick,inner sep=0pt,minimum size=10mm},
emp/.style={circle,draw=white!50,fill=white!20,thick,inner sep=0pt,minimum size=10mm},
bend angle=20, 
pre/.style={<-,shorten <=1pt,>=stealth',semithick}, 
post/.style={->,shorten >=1pt,>=stealth',semithick}]
         \node[place]  (ni)  at ( 0,0) {i};        
         \node[place]  (nj)  at ( 3,0) {j}
         	edge [->, bend right, thick] node[auto] {$B_j^i$} (ni)
          	edge [<-, bend left, thick] node[auto] {$B_i^j$}(ni);
	\node[emp] (nni) at (-2,0){}
		edge[->, bend right, thick] node[auto]{...}(ni)
		edge[<-, bend left, thick] node[auto]{}(ni);
	\node[emp] (nnj) at (5,0){}
		edge[->, bend right, thick] node[auto]{...}(nj)
		edge[<-, bend left, thick] node[auto]{}(nj);		
	\node[emp] (nniu) at (0,2){}
		edge[->, bend right, thick] node[auto]{...}(ni)
		edge[<-, bend left, thick] node[auto]{}(ni);
	\node[emp] (nnid) at (0,-2){}
		edge[->, bend right, thick] node[auto]{...}(ni)
		edge[<-, bend left, thick] node[auto]{}(ni);
	\node[emp] (nnju) at (3,2){}
		edge[->, bend right, thick] node[auto]{...}(nj)
		edge[<-, bend left, thick] node[auto]{}(nj);
	\node[emp] (nnjd) at (3,-2){}
		edge[->, bend right, thick] node[auto]{...}(nj)
		edge[<-, bend left, thick] node[auto]{}(nj);		
\end{tikzpicture}
\caption{A schematic representation of an open quantum walk on a graph with nodes $i$ and $j$. The operators $B_i^j$ represent the transition operators of the walker.}
\end{center}
\label{fig:Figure1}       
\end{figure}
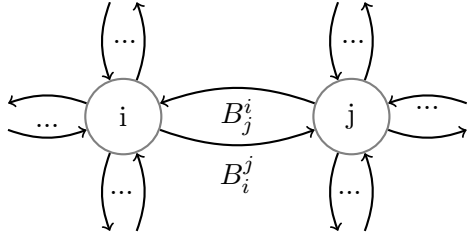

To describe the dynamics of the internal degree of freedom of the quantum walker for each edge $(i,j)$ we introduce  a bounded transition operator $B^i_j\in\cal{H}$. This operator describes the change in the internal degree of freedom of the walker due to the "quantum jump" from node $j$ to node $i$ (see Fig. 1). By imposing for each $j$ that, 
\begin{equation}\label{eq1}
\sum_i {B^i_j}^\dag B^i_j= I,
\end{equation}
we insure, that for each node of the graph $j\in\mathcal{V}$ there is a corresponding trace preserving completely positive map on the operators of $\mathcal{H}$:
\begin{equation}
\mathcal{M}_j(\tau)=\sum_i B^i_j \tau {B^i_j}^\dag.
\end{equation}

The transition operators $B^i_j$ act only on the ``quantum coin" Hilbert space $\mathcal{H}$ and do not perform transitions from node $j$ to node $i$, they can be easy extended to operators $M^i_j\in\mathcal{H}\otimes\mathcal{K}$ acting on total Hilbert space in the following way
\begin{equation}
M^i_j=B^i_j\otimes | i\rangle\langle j|\,.
\end{equation}
It is clear that, if the condition expressed in Eq. (\ref{eq1}) is satisfied, then $\sum_{i,j} {M^i_j}^\dag M^i_j=1$. This condition defines a trace preserving completely positive map for  density matrices on $\mathcal{H}\otimes\mathcal{K}$, i.e.,
\begin{equation}\label{OQW}
\mathcal{M}(\rho)=\sum_i\sum_j M^i_j\,\rho\, {M^i_j}^\dag.
\end{equation}
This  map defines the discrete time \textit{Open Quantum Walk} (OQW)  \cite{pla, JSP}.
It is clear that for an arbitrary initial state the density matrix $\sum_{i,j} \rho_{i,j}\otimes| i\rangle\langle j|$ will take a diagonal form after just one step of the open quantum  random walk Eq. (\ref{OQW}). One can see that,
\begin{eqnarray}
\nonumber
\mathcal{M}\left(\sum_{k,m} \rho_{k,m}\otimes|k\rangle\langle m|\right)&=&\sum_{i,j,k,m} B^i_j\otimes | i\rangle\langle j|\,\left(\rho_{k,m}\otimes|k\rangle\langle m|\right)\, {B^i_j}^\dag\otimes | j\rangle\langle i|\\\nonumber
&=&\sum_{i,j,k,m} B^i_j\rho_{k,m}{B^i_j}^\dag\otimes | i\rangle\langle i|\delta_{j,k}\delta_{j,m}\\
&=&\sum_i \left(\sum_j B^i_j\rho_{j,j}{B^i_j}^\dag\right)\otimes | i\rangle\langle i|.
\end{eqnarray}

Hence, in the following,  we will assume that the initial state of the system has the form
\begin{equation}\label{rho}
\rho=\sum_i \rho_i\otimes | i\rangle\langle i|,
\end{equation}
with
\begin{equation}
\sum_i \mathrm{Tr}(\rho_i)=1.
\end{equation}
It is easy to give an explicit iteration formula for the OQW from step $n$ to step $n+1$ 
\begin{equation}\label{rho}
\rho^{[n+1]}=\mathcal{M}(\rho^{[n]})=\sum_i \rho_i^{[n+1]}\otimes | i\rangle\langle i|,
\end{equation}
where
\begin{equation}
\rho_i^{[n+1]}=\sum_j B^i_j \rho_j^{[n]}{B^i_j}^\dag.
\end{equation}

The above iteration formula gives a clear physical meaning to the mapping that we introduced: the state of the system on  site $i$ is determined by the conditional shift from all connected sites $j$, which are defined by the explicit form of the open quantum coin operators $B_j^i$. Also, one can see that $\mathrm{Tr}[\rho^{[n+1]}]=\sum_i\mathrm{Tr}[\rho_{i}^{[n+1]}]=1$.

\section{Examples of OQW}
\subsection{OQW on a 2-node graph}
As an example of open quantum walk we consider the simplest case of a walk on a 2-node graph (see Fig. 2a). In this case the transition operators $B_i^j$ $(i,j=1,2)$ satisfy:
\begin{equation}
B_1^{1\dag}B_1^1+B_1^{2\dag}B_1^2=I,\quad B_2^{2\dag}B_2^2+B_2^{1\dag}B_2^1=I.
\end{equation}
The state of the walker $\rho^{[n]}$ after $n$ steps is given by,
\begin{equation}
\rho^{[n]}=\rho_1^{[n]}\otimes|1\rangle\langle 1|+\rho_2^{[n]}\otimes|2\rangle\langle 2|,
\label{2node}
\end{equation}
where the particular form of the $\rho_i^{[n]}$ $(i=1,2)$ is found by recursion,
\begin{eqnarray}
\rho_1^{[n]}=B_1^1\rho_1^{[n-1]}B_1^{1\dag}+B_2^1\rho_2^{[n-1]}B_2^{1\dag},\\\nonumber
\rho_2^{[n]}=B_1^2\rho_1^{[n-1]}B_1^{2\dag}+B_2^2\rho_2^{[n-1]}B_2^{2\dag}.
\end{eqnarray}

\begin{figure}
\begin{center}
\begin{tikzpicture}
[place/.style={circle,draw=black!50,fill=white!20,thick,inner sep=0pt,minimum size=10mm},
emp/.style={circle,draw=white!50,fill=white!20,thick,inner sep=0pt,minimum size=10mm},
bend angle=25, 
pre/.style={<-,shorten <=1pt,>=stealth',semithick}, 
post/.style={->,shorten >=1pt,>=stealth',semithick}]
         \node[place]  (n0)  at  (0,0)  {\textcolor{black}{$1$}}
         	edge[loop left, thick] node [auto] {$B_1^1$}(n0);
         \node[place]  (n1)  at  (3,0)  {\textcolor{black}{$2$}}
         	edge [->, bend right, thick] node[above] {$B_1^2$} (n0)
          	edge [<-, bend left, thick] node[auto] {$B_1^2$}(n0)
		edge[loop right, thick] node [auto] {$B_2^2$}(n0);
         \node[emp]  (e1)  at  (-2.5,0) [label=north:$\mathrm{(a)}$] {};  
\end{tikzpicture}
\\
\begin{tikzpicture}
[place/.style={circle,draw=black!50,fill=white!20,thick,inner sep=0pt,minimum size=6mm},
emp/.style={circle,draw=white!50,fill=white!20,thick,inner sep=0pt,minimum size=6mm},
bend angle=45, 
pre/.style={<-,shorten <=1pt,>=stealth',semithick}, 
post/.style={->,shorten >=1pt,>=stealth',semithick}]
         \node[place]  (n0)  at  (2.4,0)  {\textcolor{black}{$0$}};
         \node[place]  (n1)  at  (3.6,0)  {\textcolor{black}{$+1$}}
         	edge [->, bend right, thick] node[above] {$C$} (n0)
          	edge [<-, bend left, thick] node[auto] {$B$}(n0);
         \node[place]  (n2)  at  (4.8,0)  {\textcolor{black}{$+2$}}
         	edge [->, bend right, thick] node[above] {$C$} (n1)
          	edge [<-, bend left, thick] node[auto] {$B$}(n1);
         \node[place]  (nm1)  at  (1.2,0)  {\textcolor{black}{$-1$}}
         	edge [->, bend right, thick] node[auto] {$B$} (n0)
          	edge [<-, bend left, thick] node[auto] {$C$}(n0);
         \node[place]  (nm2)  at  (0,0)  {\textcolor{black}{$-2$}}
         	edge [->, bend right, thick] node[auto] {$B$} (nm1)
          	edge [<-, bend left, thick] node[auto] {$C$}(nm1);
         \node[emp]  (e1)  at  (-1.2,0) [label=north:$\mathrm{(b)}$]   {}
         	edge [->, bend right, dotted, thick] (nm2)
		edge [<-, bend left, dotted, thick] (nm2);
         \node[emp]  (e2)  at  (6,0)  {}
         	edge [->, bend right, dotted, thick] (n2)
		edge [<-, bend left, dotted, thick] (n2);
\end{tikzpicture}
\\
$
\begin{array}{ccc} 
\includegraphics[width= .32\linewidth]{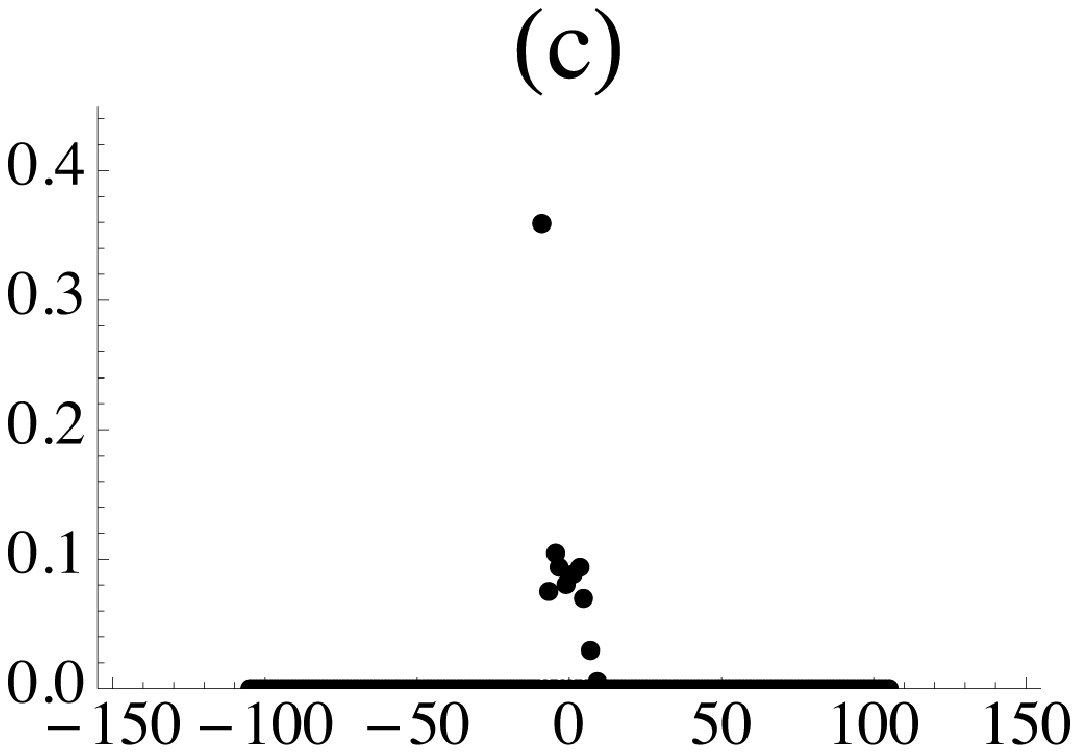} &
\includegraphics[width= .32\linewidth]{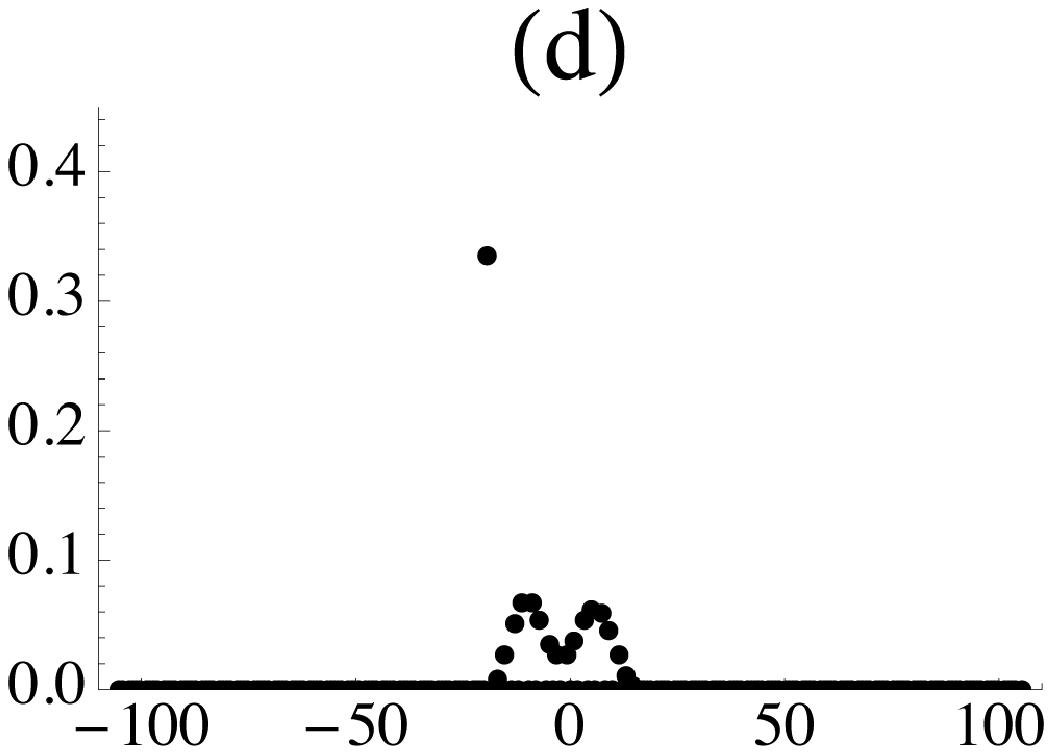} &
\includegraphics[width= .32\linewidth]{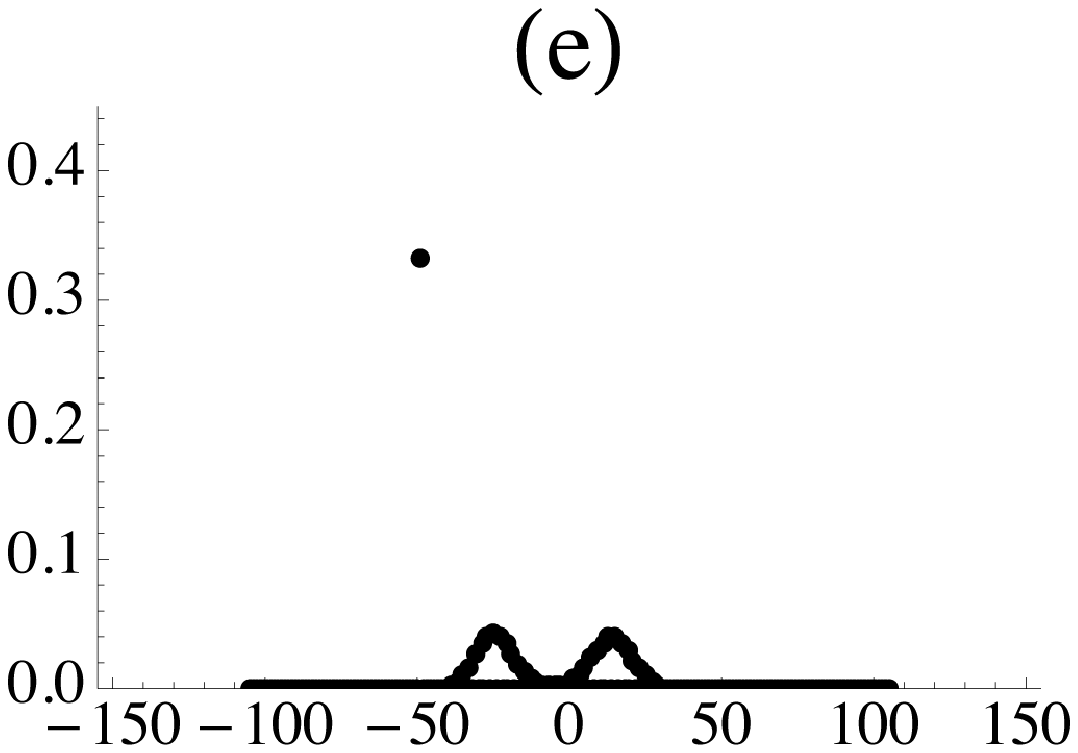}
\end{array}$
\end{center}
\caption{(a) A schematic representation of an open quantum walk on a 2-node graph. The operators $B_i^j$ $(i, j = 1,2)$ represent the transition operators of the walk. Open quantum walk on Z. (b) A schematic representation of the OQW on $\mathbb{Z}$: all transitions to the right are induced by the operator $B$, while all transitions i to the left are induced by the operator $C$; Figures (c)-(e) show the 
occupation probability distribution for the "walker" with the initial state $\rho^{[0]}=\frac{1}{3}I_3\otimes |0\rangle\langle 0|$ and transition operators given by Eq. (\ref{eq:BCpart}) after 10, 20 and 50 steps, respectively.}
\label{fig:OQW2}
\end{figure}

\subsection{OQW on $\mathbb{Z}$}
In this subsection we consider the case of OQWs on $\mathbb{Z}$ with transition between neighbouring nodes (see Fig. 2b). In this case, the general expression for the open quantum walk on the graph reads,
\begin{equation}\label{rhoz}
\rho^{[n+1]}=\mathcal{M}(\rho^{[n]})=\sum_i \rho_i^{[n+1]}\otimes | i\rangle\langle i|,
\end{equation}
where
\begin{equation}
\rho_i^{[n+1]}=B^i_{i+1} \rho_{i+1}^{[n]}{B^{i\dag}_{i+1}}+B^i_{i-1} \rho_{i-1}^{[n]}{B^{i\dag}_{i-1}}.
\end{equation}
Here we consider a homogenous open quantum walk (which implies that $\forall i,B_i^{i+1}\equiv B$ and $B_i^{i-1}\equiv C$).
For an initially localised walker in the node 0, namely
\begin{equation}
\rho^{[0]}=\rho^{[0]}_0 \otimes |0\rangle\langle 0|,
\end{equation}
the density matrix after one  $\rho^{[1]}$ and two steps $\rho^{[2]}$ reads,
\begin{equation}
\rho^{[1]}=B \rho^{[0]}_0 B^\dag \otimes |1\rangle\langle 1|+C \rho^{[0]}_0 C^\dag\otimes |-1\rangle\langle -1|,
\end{equation}
and
\begin{equation}
\rho^{[2]}=B \rho^{[1]}_1 B^\dag \otimes |2\rangle\langle 2|+(C\rho^{[1]}_1 C^\dag+B\rho^{[1]}_{-1} B^\dag)\otimes |0\rangle\langle 0|+C \rho^{[1]}_{-1} C^\dag\otimes |-2\rangle\langle -2|.
\end{equation}
In order to demonstrate the dynamics of the probability distribution, we choose $B$ and $C$ in the following way,
\begin{equation}
B=\left(\begin{array}{ccc} 1 & 0 & 0 \\ 0 & \sqrt{3}/2 & 0 \\ 0 & 0 & 3/5
\end{array}\right),\quad C=\left(\begin{array}{ccc} 0 & 0 & 0 \\ 0 & 1/2 & 0 \\ 0 & 0 & 4/5
\end{array}\right).
\label{eq:BCpart}
\end{equation}
If we choose the initial state of the walker to be localised in the node $0$ with unpolarised state of an internal degree of freedom, i.e.,
$\rho^{[0]}=\frac{1}{3}I_3\otimes |0\rangle\langle 0|$, then by iteration we obtain the probability distribution of the walker after an arbitrary number of steps. In Fig. 2(c)-(e) we show the probability to find a "walker" on a particular lattice site for different numbers of steps. Already, after $20$ steps (Fig. 2(d)) one can see the formation of a ``soliton"- like distribution and two gaussian packets moving in different directions. After $50$ steps (Fig. 2(e)) this behaviour is even more certain.

\begin{figure}[h]
$
\begin{array}{cc} 
\includegraphics[width= .48\linewidth]{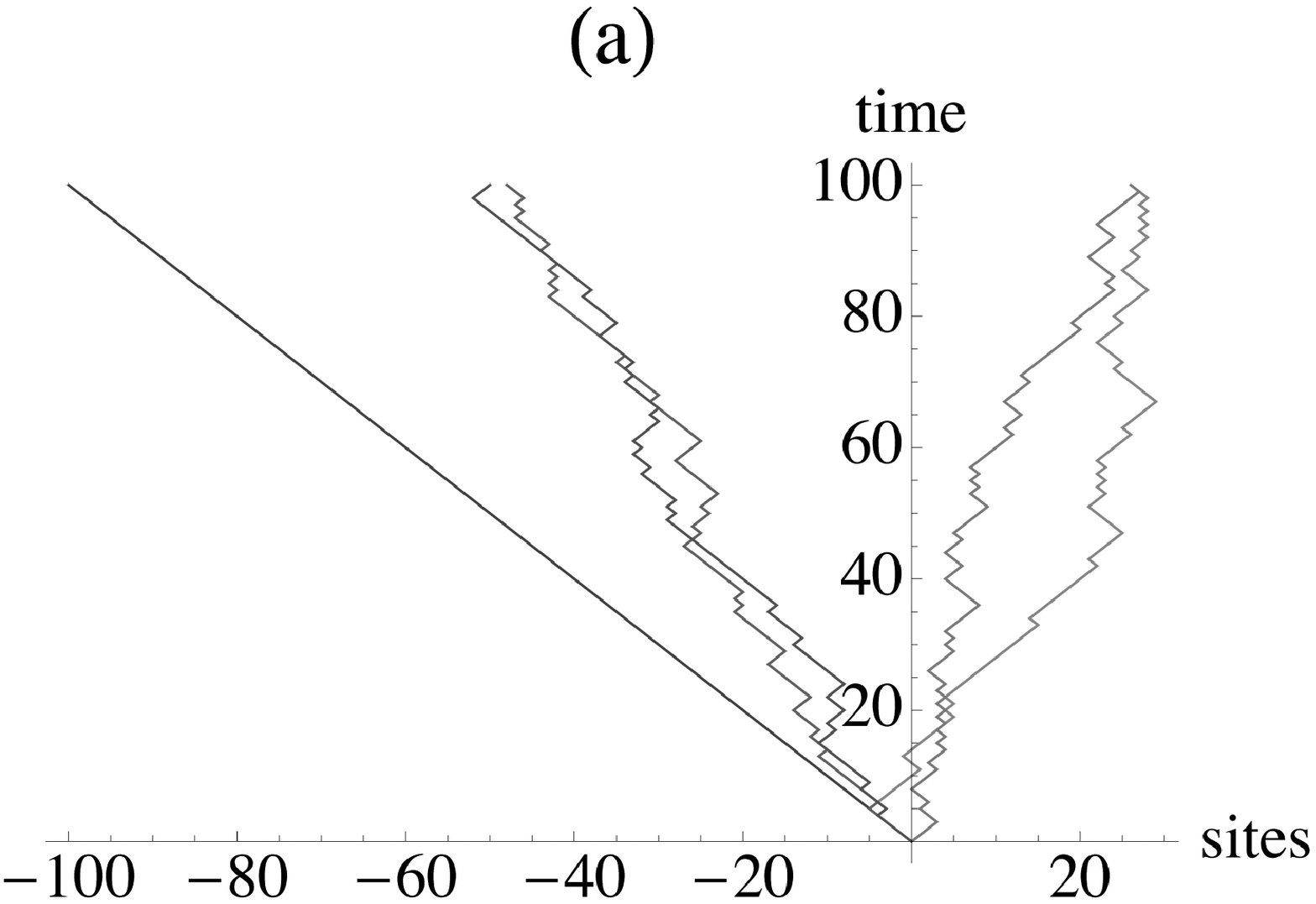} &
\includegraphics[width= .48\linewidth]{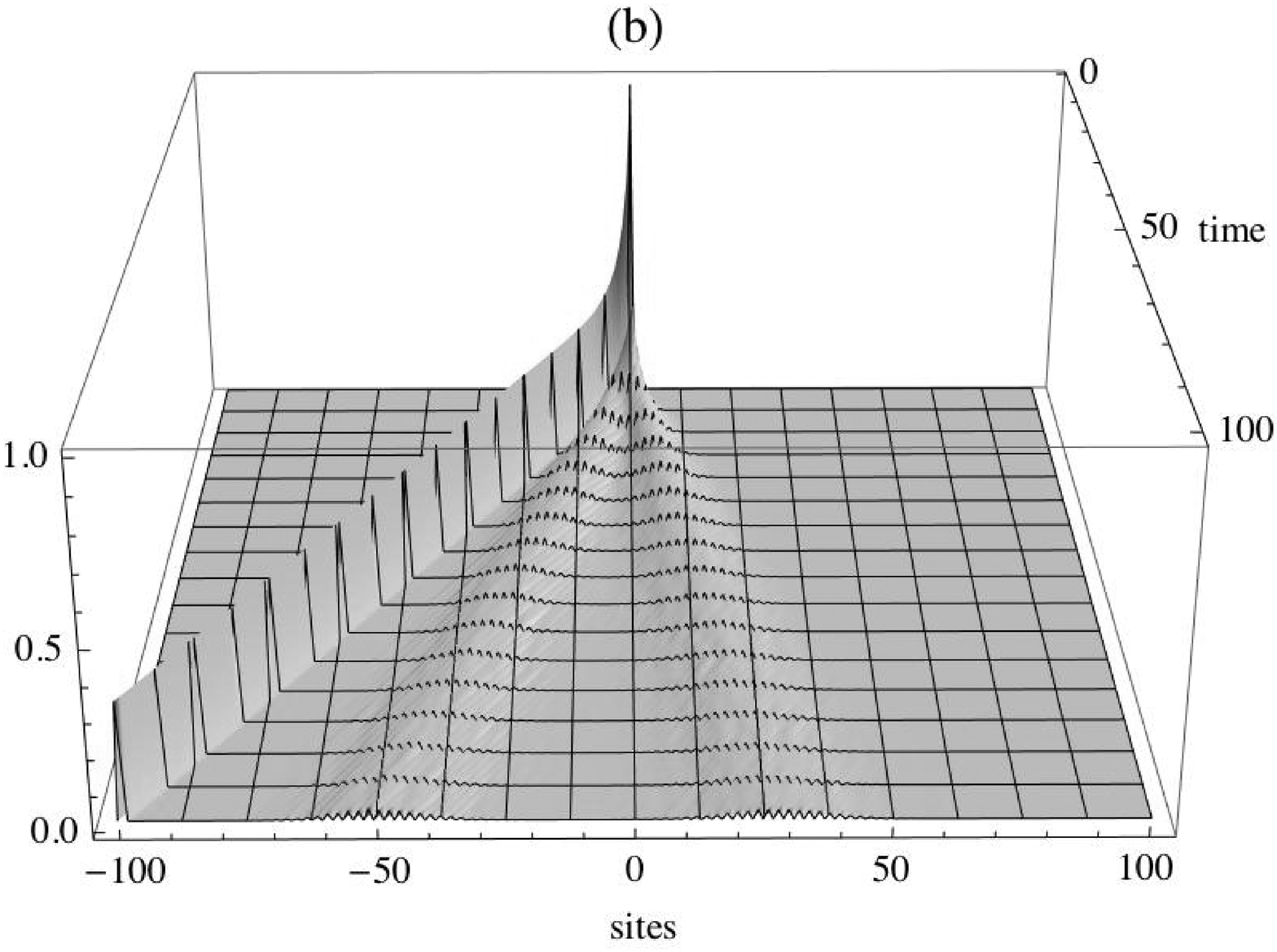}
\end{array}$
\caption{Simulation of the OQW in term of quantum trajectories. Examples of trajectories are shown in (a); (b) shows the average over $10^5$ trajectories}
\end{figure}

\section{Unravelling of OQW}

It is interesting to mention that OQW can be simulated by the means of \emph{Quantum Trajectories} \cite{JSP}. To introduce this formalism we start by considering a particular case of initial state of the system, namely, a walker that is localized at a single node,
\begin{equation}
\rho^{[0]}=\rho\otimes | i\rangle\langle i|.
\end{equation}
After one step of OQW the state of the walker is given by
\begin{equation}
\rho^{[1]}=\sum_j (B_i^j \rho {B_i^j}^\dag)\otimes |j\rangle\langle j|\,.
\end{equation}
The probability to find the walker at the site $j$ is given by $p_j=\tr(B_i^j \rho {B_i^j}^\dag)$. If one performs measurements of the position of the walker at node $j$ the state of the walker reads
\begin{equation}
\frac 1{p_j} (B_i^j \rho {B_i^j}^\dag)\otimes| j\rangle\langle j|\,.
\end{equation}
Repetition of this procedure gives rise to a classical Markov chain, valued in the set of states of the form $\rho\otimes |i\rangle\langle i|\,$.
One can see that this procedure on average will simulate a master equation driven by $\mathcal{M}$:
\begin{equation}
\mathcal{E}[\rho_{n+1}]=\sum_j p_j\frac 1{p_j} (B_i^j \rho_n {B_i^j}^\dag)\otimes|j\rangle\langle j|=\sum_j  (B_i^j \rho_n {B_i^j}^\dag)\otimes |j\rangle\langle j|=\mathcal{M}(\rho_n)\,.
\end{equation}

If one choose the initial state of the system in the pure state $\rho=|\phi\rangle\langle\phi|\otimes | i\rangle\langle i|$, then the realization remains in a pure state. It is easy to see that, an arbitrary initial pure state $| \phi\rangle \otimes| i\rangle$ will jump randomly to one of the states 
\begin{equation}
\frac 1{\sqrt{p^j_i}}\,B^j_i|  \phi\rangle\otimes | j\rangle
\end{equation}
with probability
\begin{equation}
p^j_i=||B^j_i |  \phi\rangle||^2\,.
\end{equation}

This procedure leads to a classical Markov chain valued in the space of wave functions of the form $|\phi\rangle\otimes | i\rangle$. On average, this random walk simulates the OQW driven by $\cal{M}$. 

As an example of this unravelling we consider an OQW on a line. The transition matrices $B$ and $C$ are given by the Eq. (\ref{eq:BCpart})
and the initial state of the walker is
$$
|\psi_0\rangle=\frac{1}{\sqrt{3}}(|1\rangle+|2\rangle+|3\rangle)\otimes|0\rangle.
$$
The results of the simulations are presented in Fig. 3. In Fig. 3(a) we show five different quantum trajectories, one can clearly see the different qualitative character of some of them. The average over $10^5$ trajectories is shown in Fig. 3(b).

\begin{figure}
\begin{center}
\begin{tikzpicture}
[place/.style={circle,draw=blue!50,fill=blue!20,thick,inner sep=0pt,minimum size=10mm},
emp/.style={rectangle,draw=white!50,fill=white!20,thick,inner sep=0pt,minimum size=10mm},
bend angle=20, 
pre/.style={<-,shorten <=1pt,>=stealth',semithick}, 
post/.style={->,shorten >=1pt,>=stealth',semithick}]

         \node[emp]  (rn)   at  (0,4) {$\rho^{[n]}$}; 
         \node[emp]  (rs1) at  (0,2) {$\rho^{\mathrm{ext}}\in\mathcal{H}\otimes\mathcal{K}\otimes\mathcal{K}$}
         	edge[<-] node[auto] {Step 1}(rn);
	\node[emp]   (rs2) at (0,0) {$\cal{U}\rho^{\mathrm{ext}}\cal{U}^\dag$}
		edge[<-] node[auto] {Step 2}(rs1);
	\node[emp]   (rs3) at (4,0) {$\{P_k=|k\rangle\langle k|\}\in\mathcal{K}$}
		edge[<-] node[auto] {Step 3}(rs2);
	\node[emp]  (rs4) at  (8,0) {swap in $\mathcal{K}\otimes\mathcal{K}$}
		edge[<-] node[auto] {Step 4}(rs3);
	\node[emp]  (rs5) at  (8,2) {$\mathrm{Tr_{\mathcal{K}}[\tilde{\rho}^{\mathrm{ext}}]}$}
		edge[<-] node[auto] {Step 5}(rs4);
	\node[emp]  (f) at  (8,4) {$\rho^{[n+1]}=\mathcal{M}(\rho^{[n]})$}
		edge[<-] (rs5)
		edge[<-] node[auto] {OQW} (rn);
		
\end{tikzpicture}
\end{center}
\caption{Scheme of the various steps required for the \emph{realisation procedure} of the Open Quantum Walk.}
\label{fig:PhysicalRealization}
\end{figure}
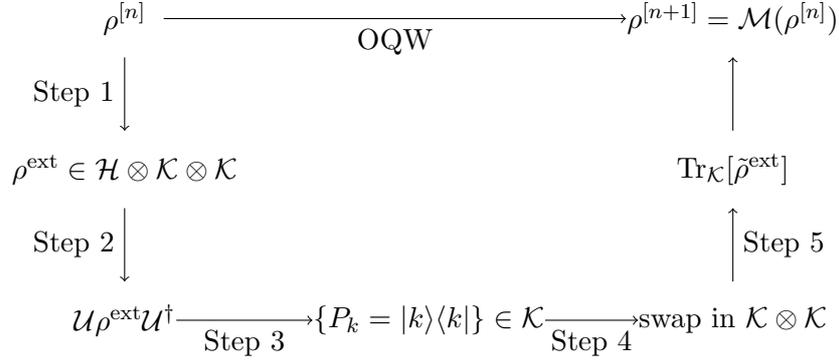

\section{Realisation procedure} 

It is natural to ask how to obtain an OQW by reduction of the unitary dynamics of an appropriately enlarged system. The situation is depicted in 
Fig. \ref{fig:PhysicalRealization} and will be described step by step \cite{JSP}. To achieve this goal we need to extend the Hilbert space of the system by an ancilla-graph identical to the graph described by the space $\cal{K}$ and define on this extended Hilbert space $\mathcal{H}\otimes\mathcal{K}\otimes\mathcal{K}$ a specially chosen unitary operator $\mathcal{U}$. It is always possible for any given $k$  to define a unitary operator $U(k)\in\mathcal{H}\otimes\mathcal{K}$,
such that the first column of the operator is given by $B^i_k$ and by filling the rest with corresponding elements to fulfil the unitary condition 
\begin{equation}
U^i_1(k)=B^i_k,\qquad\sum_j U^{i'}_j(k)^\dag U^i_j(k)=\delta_{i,i'} I.
\end{equation}
The unitary operator $\mathcal{U}$ is a diagonal sum of the operators $U(k)$ with corresponding projector representation
\begin{equation}
\mathcal{U}=\sum_k U(k)\otimes|k\rangle\langle k|.
\end{equation}
Again, we assume the initial state $\rho=\sum_i\rho_i\otimes|i\rangle\langle i|\in\mathcal{H}\otimes\mathcal{K}$. 

\textit{The first step} is to extend the state into a larger Hilbert space in the following way
\begin{equation}
\rho=\sum_i\rho_i\otimes|i\rangle\langle i|\rightarrow\rho^{\mathrm{ext}}=\sum_i\rho_i\otimes|1\rangle\langle 1|\otimes|i\rangle\langle i|,
\end{equation}
so that the extended density matrix $\rho^{\mathrm{ext}}\in\mathcal{H}\otimes\mathcal{K}\otimes\mathcal{K}$.

\textit{Step 2} is the application of the unitary operator $\mathcal{U}$ to the extended density matrix $\rho^{\mathrm{ext}}$, i.e., 
\begin{equation}
\mathcal{U}\rho^{\mathrm{ext}}\mathcal{U}^\dag=\sum_{i,j,p}B_i^j\rho_i B_i^{p\dag}\otimes |j\rangle\langle p|\otimes |i\rangle\langle i|.
\end{equation}

In \textit{Step 3} we perform a full set of measurements $P_k=I_{\mathcal{H}}\otimes|k\rangle\langle k|\otimes I_{\mathcal{K}}$ on the "extra" position space or we 
subject the system  to decoherence in this "extra" position subspace. As result of Step 3 we obtain the following density matrix
\begin{equation}
\sum_{i,j}B_i^j\rho_i B_i^{j\dag}\otimes |j\rangle\langle j|\otimes |i\rangle\langle i|.
\end{equation}

\textit{In Step 4} we swap  position subspaces. The result of this operation is the following density matrix
\begin{equation}
\sum_{i,j}B_i^j\rho_i B_i^{j\dag}\otimes |i\rangle\langle i|\otimes |j\rangle\langle j|.
\end{equation}

\textit{The last step} (Step 5) is  tracing out the "extra" position subspace and one obtains the OQW as promised
\begin{equation}
\sum_{i,j}B_i^j\rho_i B_i^{j\dag}\otimes |j\rangle\langle j|.
\end{equation}

\section{Connection to Unitary Quantum Walk and Classical Random Walk}
\subsection{Connection to unitary random walk}

The \textit{unitary quantum walk} (UQW) can be recovered from the OQW realisation procedure by a special choice of the  transition operators $B_i^j$ and by excluding from the realisation procedure illustrated in Fig. \ref{fig:PhysicalRealization} Step 3 (measurements or decoherence in position space). This is easily demonstrated for a UQW  on $\mathbb{Z}$. In this case of a UQW on the line an additional condition needs to be imposed on the transition operators $B$ and $C$: next to the normalisation condition $B^\dag B+C^\dag C=I$ we need to request that  $C^\dag B=0$. In this case the sum of $B$ and $C$ is a unitary operator. Explicitly, in order to recover  the Hadamard UQW on the line we may choose  
\begin{equation}
B=\left(\begin{array}{cc} \alpha & \beta \\ 0 & 0
\end{array}\right),\quad C=\pm\left(\begin{array}{cc} 0 & 0 \\ -\beta^* & \alpha^*\end{array}\right),
\end{equation}
where $|\alpha|^2+|\beta|^2=1$ and $\alpha,\beta\in\mathbb{C}$.
Starting from the state $|\psi_0\rangle\otimes|0\rangle$ and following the  realization procedure (Fig. \ref{fig:PhysicalRealization}) while skipping  Step 3,  the state of the system after one step is given by $B|\psi_0\rangle\otimes|+1\rangle+C|\psi_0\rangle\otimes|-1\rangle$. This is exactly a Hadamard UQW on $\mathbb{Z}$. The general condition for obtaining UQWs from OQWs can be found in \cite{JSP}.

\subsection{The classical random walk}
It is interesting to note that  the OQWs contains, as a special case, the  classical random walk as well. To see this, we consider the case $\cal{H}=\cal{K}=\mathbb{C}^V$, and we introduce a matrix $P=\{P_{i,j}\}$ of classical transition probabilities on the graph $\cal{V}$ with standard normalisation condition $\sum_iP_{j,i}=1$ and add an arbitrary family of unitary operators, $U_i^j\in\mathbb{C}^\mathcal{V}$. In this case, if we choose the transition operators $B_i^j$ to have the form $B_i^j=\sqrt{P_{i,j}}U_i^j$. Then, for the initial state $\rho=\sum_k\rho_k\otimes|k\rangle\langle k|$ the probability to find a ``walker" after one step on the site $i$ reads,
\begin{equation}
p^{[1]}(i)=\mathrm{Tr}(\mathcal{M}(\rho)|i\rangle\langle i|)=\sum_k P_{k,i}\mathrm{Tr}(\rho_k).
\end{equation} 
After two steps the probability to find walker on the site $i$ will be given by
\begin{equation}
p^{[2]}(i)=\mathrm{Tr}(\mathcal{M}(\mathcal{M}(\rho))|i\rangle\langle i|)=\sum_{k,m}p_{m,k}p_{k,i}\mathrm{Tr}(\rho_m).
\end{equation}
Thus, the probability of transition does not depend on the internal degrees of freedom of the ``walker". It only  depends on the classical transition probability matrix $P$, as expected for a classical random walk. 

\section{Limit theorems for OQWs}

Recently Attal \textit{et al} \cite{clt1} and Konno and Yoo \cite{clt2} have formulated central limit theorems for OQWs. In particular, Attal \textit{et al} have shown that if we have an OQW on $\mathbb{Z}^d$ and $\rho_\infty$ is a unique invariant state of the OQW, then the normalised quantum trajectory of the walk will converge to the Gaussian distribution $\mathcal{N}(0,C)$ in $\mathbb{R}^d$ with covariance matrix $C$. Similar results were obtained by Konno and Yoo for OQW on $\mathbb{Z}$. Sinayskiy and Petruccione \cite{PhysSc} have analysed the limit distribution of the OQW on $\mathbb{Z}$ for the case of simultaneously diagonalisable transition operators. They found that the number of Gaussian distributions in the limiting case is bounded by the dimension of the transition operator and defined by its spectrum.

The asymptotic behaviour of the example of OQW on $\mathbb{Z}$ considered in this proceedings fits into the case studied by Sinayskiy and Petruccione in Ref. \cite{PhysSc}. The transition operators $B$ and $C$ Eq. (\ref{eq:BCpart}) are simultaneously diagonalizable. The dimension of the tradition operators is 3 which implies maximally three different asymptotic distributions (see Fig.2(e)). The condition for a ``soliton" like distribution is the presence of zeros in spectrum of one of the transition operators. In this case transition operator $C$ has a zero eigenvalue. Two other non-zero eigenvalues are describing the velocity of drift and the width of two Gaussian distributions.

\section{Quantum Computing application}
Recently, Verstraete \textit{et al.} \cite{dqc} proposed a dissipative model of quantum computing (DQC), capable of performing universal quantum computation.
The dissipative quantum computing setup consists of a linear chain of time registers. Initially, the system is in a time register labeled by $0$. The result of the computation is measured in the last time register labeled by $T$. Neighboring time registers are coupled to local baths. The result of the computation can be read out from the time-register $T$. In particular, for a quantum circuit given by the set of unitary operators $\{ U_t\}_{t=1}^T$ the final state of the system is given by $|\psi_T\rangle=U_TU_{T-1}\ldots U_2U_1|\psi_0\rangle$. It is shown that the unique steady state of the system in this case will be 
\begin{equation}
\rho=\frac{1}{T+1}\sum_t|\psi_t\rangle\langle \psi_t|\otimes|t\rangle\langle t |.
\end{equation}
Recently, it was shown that, using the formalism of OQWs one can perform dissipative quantum computations with higher efficiency than the original scheme of DQC. The explicit  open quantum walk implementation of the Toffoli gate and the Quantum Fourier Transform has been reported \cite{QIP}. If one needs to implement a circuit with a set of unitaries so that initial and final states are connected as $|\psi_T\rangle=U_TU_{T-1}\ldots U_2U_1|\psi_0\rangle$, then the corresponding OQW implementation will be given by a linear chain of $T+1$ nodes with corresponding iteration formula:
\begin{equation}
\rho_j^{[n]}=\omega U_j\rho_{j-1}^{[n-1]}U_j^\dag+\lambda U_{j+1}^\dag\rho_{j+1}^{[n-1]}U_{j+1}, j=1\ldots (T-1),
\end{equation}
\begin{equation}
\rho_0^{[n]}=\lambda \rho_{0}^{[n-1]}+\lambda U_{1}^\dag\rho_{1}^{[n-1]}U_{1},
\end{equation}
\begin{equation}
\rho_T^{[n]}=\omega \rho_{T}^{[n-1]}+\omega U_{T-1}\rho_{T-1}^{[n-1]}U_{T-1}^\dag,
\end{equation}
where the constants $\omega$ and $\lambda$ are positive constants such that $\omega+\lambda=1$.

In this case the unique steady state of the OQW will have the following form,
\begin{equation}
\rho_{SS}=\sum_{i=0}^Tp_i |\psi_i\rangle\langle \psi_i|,
\end{equation}
where $p_i$ denote the steady state probabilities of detecting the walker at the node $i$. In the case of $\omega=\lambda$ all probabilities of detection will be the same, namely $p_i=1/(T+1)$. In the case $\omega>\lambda$, the probability of detection of the walker at site $T$ will be bounded between $1/(T+1)<p_T<1$.

As an example of application of OQW implementation of dissipative quantum computing model we consider a phase estimation algorithm with an unknown unitary operator acting on 4-qubit Hilbert Space \cite{NC}.
The quantum phase estimation algorithm is implemented through a sequence of Hadamard operations (1 unitary), a conditional unknown unitary gate on each ancilla qubit (4 unitaries) and a 4-qubit inverse Quantum Fourier Transform (15 unitaries). In summary, the 4-qubit phase estimation algorithm is implemented with help of 21 unitary operation. This means that an equivalent OQW scheme will contains 22 nodes (from node $0$ for initial conditions to node $21$ for the results of computation). Fig. 5(a) shows the dependence of the probability of successful performance of the phase estimation as a function of the number of steps of the walk. Curves (1)-(4) in Fig. 5(a) correspond to different values of the parameter $\omega=0.5, 0.6, 0.8, 0.9$, respectively. Curve (1) corresponds to the case $\omega=0.5$ which is the conventional dissipative quantum computing model. In Fig. 5(b) we analyse the necessary number of steps to reach the steady state and the success probability of measurement as a function of the parameter $\omega$. We observe that with increasing $\omega$ the number of steps to reach the steady state is decreasing and the probability of successful detection is increasing.

\begin{figure}
\begin{center}
\includegraphics[width=0.46\textwidth]{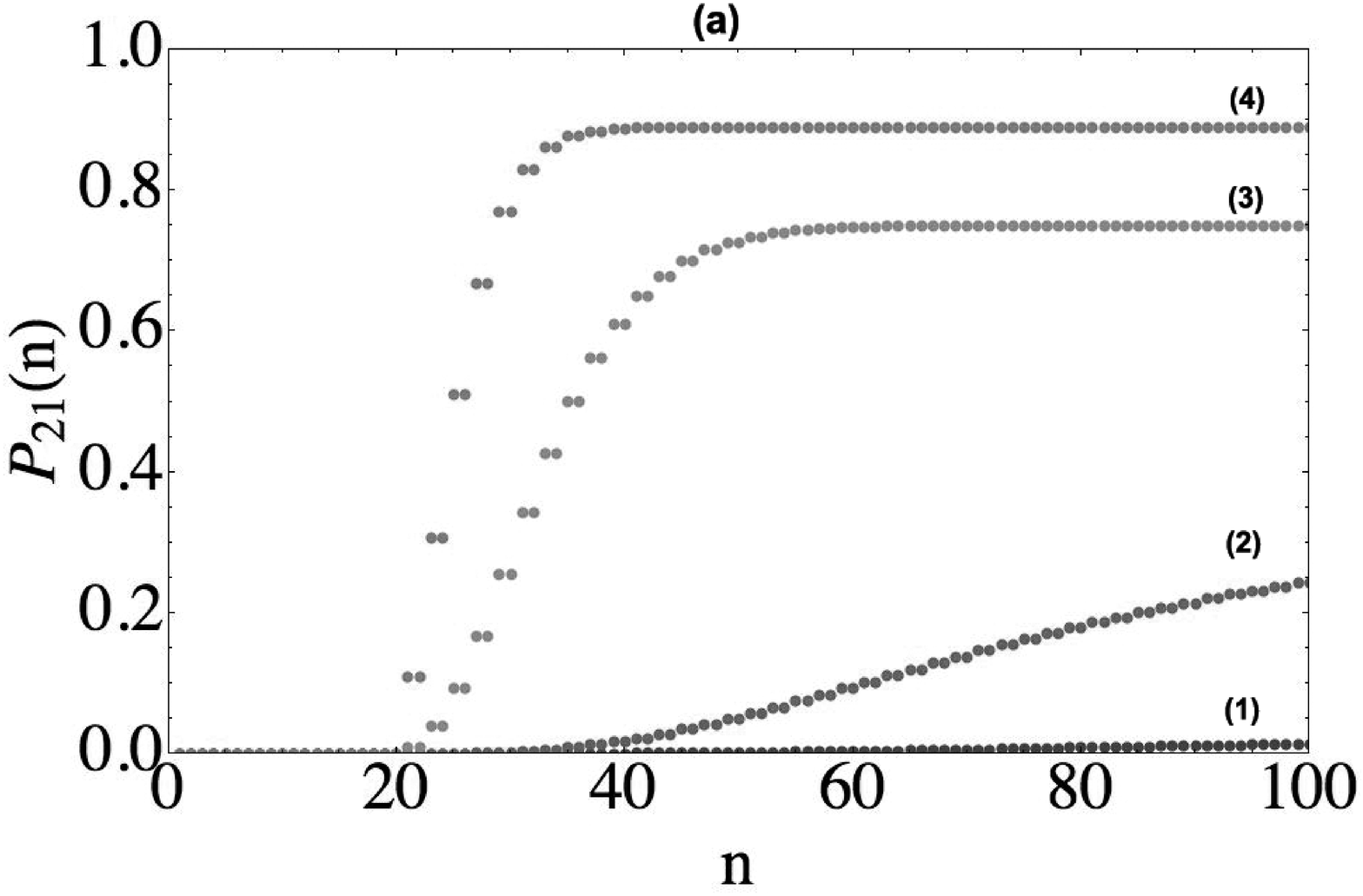}
\includegraphics[width=0.52\textwidth]{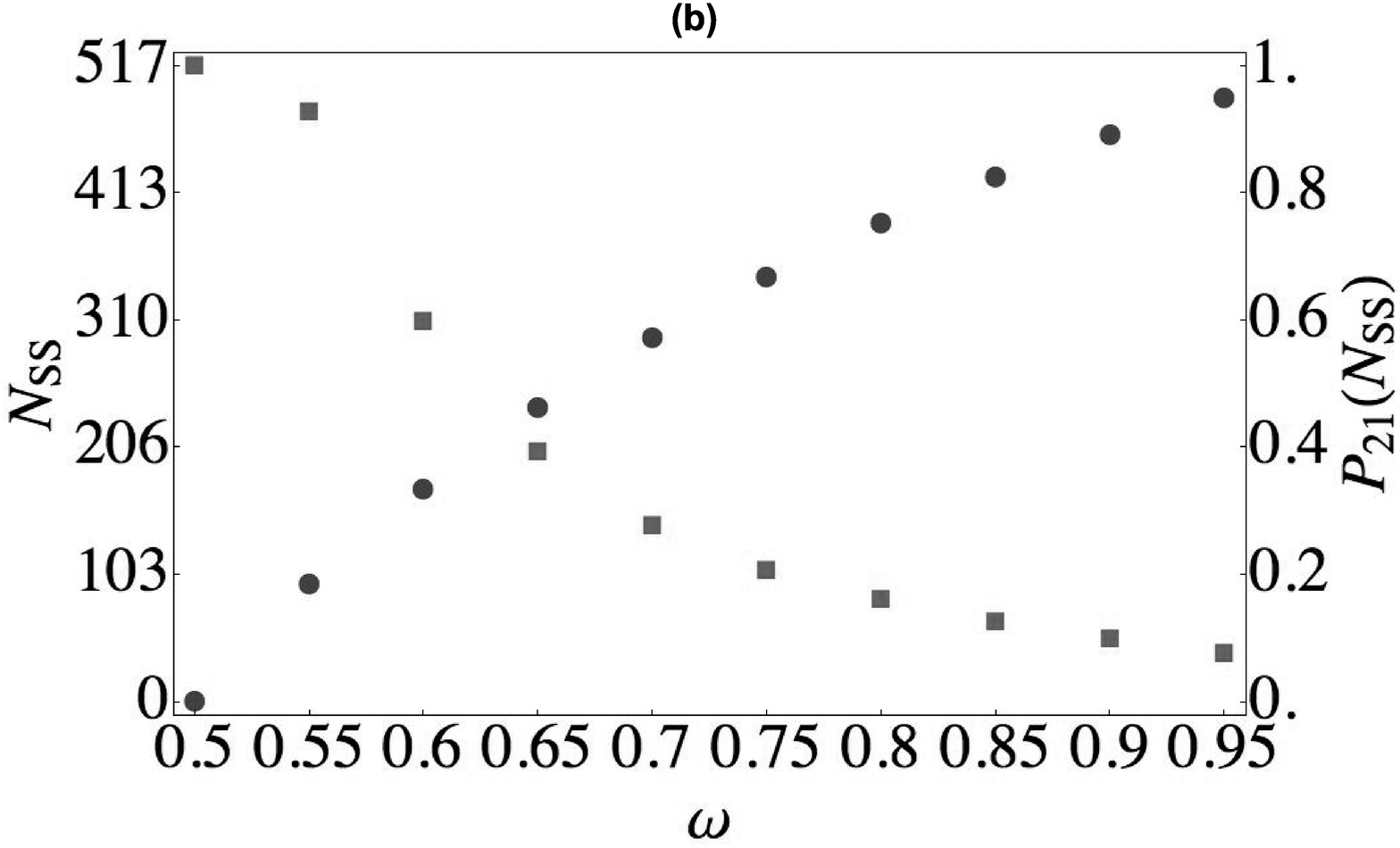}
\caption{Simulation of the efficiency of the OQW implementation of 4-qubit phase estimation algorithm; (a) shows the dynamics of the detection probability in the final node $21$ as function of the number of steps of the OQW. Curves (a1) to (a4) correspond to different values of the parameter $\omega=0.5, 0.6, 0.8, 0.9$, respectively; (b) shows the number of steps needed  to reach the steady state (squares) and the probability of detection of the successful implementation of the quantum algorithm (circles) as function of the parameter $\omega$. The number of steps to reach a steady states is simulated with $10^{-6}$ accuracy.}
\end{center}
\label{figU2}       
\end{figure}

\section{Conclusion}
In this paper we review the concept of Open Quantum Walks. We briefly present the formalism, give an example of the OQW on $\mathbb{Z}$ and apply unravelling procedure to this example. We simulate a particular case of an open quantum walk by means of quantum trajectories, give an example of quantum trajectories and perform an average over $10^5$ realisations. We summaries \textit{realisation procedure} and demonstrate a connection of OQWs to unitary quantum walks and classical random walks. We briefly mention the results of central limit theorems and show the OQW implementation of the dissipative quantum computing. As a particular example we present OQW implementation of the 4-qubit phase estimation algorithm.

This work is based upon research supported by the South African
Research Chair Initiative of the Department of Science and
Technology and National Research Foundation.

\end{document}